# Advanced Radio Frequency Timing AppaRATus (ARARAT) Technique and Applications

**Ani Aprahamian, Amur Margaryan\*, Vanik Kakoyan, Simon Zhamkochyan, Sergey Abrahamyan, Hayk Elbakyan, Samvel Mayilyan, Arpine Piloyan, Henrik Vardanyan, Hamlet Zohrabyan, Lekdar Gevorgian, Robert Ayvazyan, Artashes Papyan, Garnik Ayvazyan, Arsen Ghalumyan, Narek Margaryan, Hasmik Rostomyan, Anna Safaryan**

*A.I. Alikhanyan National Science Laboratory (Yerevan Physics Institute), Yerevan, Armenia*

**Bagrat Grigoryan, Ashot Vardanyan, Arsham Yeremyan**

*CANDLE Synchrotron Research Institute, Yerevan, Armenia*

**John Annand, Kenneth Livingston, Rachel Montgomery**

*School of Physics & Astronomy, University of Glasgow, G12 8QQ Scotland, UK*

**Patrick Achenbach, Josef Pochodzalla**

*Institut für Kernphysik, Johannes Gutenberg-Universität Mainz, Mainz, Germany*

**Dimiter L. Balabanski**

*Extreme Light Infrastructure- Nuclear Physics (ELI-NP), Bucharest-Magurele, Romania*

**Satoshi N. Nakamura**

*Department of Physics, Graduate School of Science, the University of Tokyo,Tokyo, Japan*

**Viatcheslav Sharyy, Dominique Yvon**

*Département de Physique des Particules Centre de Saclay I 91191 Gif-sur-Yvette Cedex France*

**Maxime Brodeur**

*Department of Physics and Astronomy, University of Notre Dame, Notre Dame, IN 46556, USA*

\*corresponding author: mat@mail.yerphi.am

**Abstract**

The development of the advanced Radio Frequency Timer of electrons is described. It is based on a helical deflector, which performs circular or elliptical sweeps of keV electrons, by means of 500 MHz radio frequency field. By converting a time distribution of incident electrons to a hit position distribution on a circle or ellipse, this device achieves extremely precise timing. Streak Cameras, based on similar principles, routinely operate in the ps and sub-ps time domain, but have substantial slow readout system. Here, we report a device, where the position sensor, consisting of microchannel plates and a delay-line anode, produces ~ns duration pulses which can be processed by using regular fast electronics. A photon sensor based on this technique, the Radio Frequency Photo-Multiplier Tube (RFPMT), has demonstrated a timing resolution of ~10 ps and a time stability of ~0.5 ps, FWHM. This makes the apparatus highly suited for Time Correlated Single Photon Counting which is widely used in optical microscopy and tomography of biological samples. The first application in lifetime measurements of quantum states of graphene, under construction at the A. I. Alikhanyan National Science Laboratory (AANL), is outlined. This is followed by a description of potential RFPMT applications in time-correlated Diffuse Optical Tomography, time-correlated Stimulated Emission Depletion microscopy, hybrid FRET/STED nanoscopy and Time-of-Flight Positron Emission Tomography.

**1. Introduction**

The measurement of time to very high precision is a prerequisite in many fields of science and technology. A new timing processor, the Radio Frequency Timer (RFT) [1], will be capable of ps resolution for single electron detection at high rates. Consequently, a photon sensor based on the RFT, namely the Radio Frequency Photo-Multiplier Tube (RFPMT) will be capable of detecting single photons with ps resolution at high rates. Currently there is no optical sensor capable of matching the combination of ultra-high timing resolution for single photons and very fast readout speed promised by the RFPMT, making it ideal for applications in ultra-high resolution optical microscopy.

At present, the detection of optical signals, down to the single-photon level, may be carried out with Avalanche Photodiodes (APD), vacuum Photomultiplier Tubes (PMT), Hybrid Photon Detectors (HPD) and Superconducting Nanowire Single-Photon Detectors (SNSPD). The time resolution limit of current APD, PMT or HPD for single photo-electron detection is about 100 ps full width at half maximum (FWHM), while SNSPD devices have recently reached below 5 ps [2]. The dead time of these devices is typically a few tens of ns.

By comparison the RFT, after some development, will offer around 1 ps resolution and essentially be free from dead time, so that multiple single photons resulting for example from a laser induced fluorescence could be recorded and time resolved. With fast readout from a suitably pixelated anode, the RFPMT will have enormous data throughput, potentially increasing the speed of image reconstruction by large factors.

We foresee that the RFT can offer major improvements to several imaging techniques. For example, in high-precision time-correlated, Stimulated Emission Depletion (STED) microscopy [3] precise timing offers improved coordinate resolution. Similarly in time-correlated Diffuse Optical Tomography (DOT) [4, 5] the ability of the RFPMT to map and de-convolute scattered photon time distributions with extremely high precision would be a huge advance compared to conventional photon sensors. Ultimately, with ps resolution or better, the RFPMT offers a window of opportunity to access dynamical processes in biological molecules as they take place.

The following sections describe the operational principles of the RFT (Sec.2), first measurements of the timing resolution of the RFPMT (Sec.3) and the first RFT application:

the study of quantum states with ps lifetimes (Sec.4.1). This is followed by outlines of conceptual RFPMT applications in Time-of-Flight Positron Emission Tomography (TOF-PET) (Sec.4.2), TOF-DOT (Sec. 4.3), TOF-STED (Sec.4.4), and Förster Resonance Energy Transfer (FRET) Nanoscopy (Sec. 4.5).

## 2. The Radio Frequency Timing Technique

In a typical timing technique, the time interval is measured between the leading edges of two electronic pulses applied to the start and stop inputs of a time-interval meter. A typical circuit might measure the difference in arrival time of two photons. The detectors, e.g. vacuum or Si photomultipliers, produce ~ns rise time pulses, with constant-fraction discriminators (CFD) providing sub-ns, time-pick-off precision for the logic pulses fed to time-to-digital converters (TDC).

However, timing systems based on Radio Frequency (RF) fields, the RF timing technique (RFT), can provide a resolution of order 1 ps or better. The basic principle is the conversion of information in the time domain to a spatial domain by means of a high frequency RF field [6-8]. Streak Cameras, based on similar principles, routinely operate in the ps and sub-ps time domain, but have substantial dead time associated with the readout system. In [1] we have presented a new type of RF timer, where the position sensor produces fast ns pulses, resulting in very small intrinsic dead time.

The RF timer under development by the ARARAT collaboration is shown schematically in Fig. 1 and consists of several components: UV light source, RF timing tube, RF source, position-sensing electronics, data acquisition system, power supplies and vacuum system. The RF timer is mounted in a tube, maintained at vacuum of ≤$10^{-6}$ Torr.

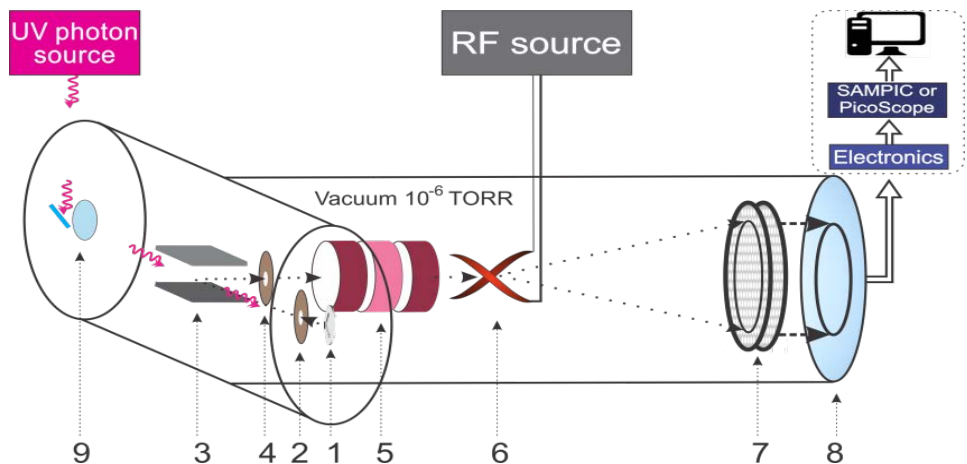

*Figure 1: Schematic of the RF timing technique. 1-Tantalum disc cathode; 2-accelerating electrode; 3-permanent magnet; 4-collimator; 5-electrostatic lens; 6-helical RF deflector; 7-dual-chevron MCP; 8-delay-line anode; 9-UV photon transmission window.*

UV photons, from a diode or pulsed laser enter the tube through a quartz window and are incident on a tantalum photocathode. Photoelectrons (PE) produced in the cathode are accelerated by a voltage V ~2.5 kV applied between the cathode (1) and an accelerating electrode (2) and then deflected through 90 deg. by a permanent magnet (3). They are collimated (4) before entering an electrostatic lens (5), which focuses the electrons on the position-sensitive detector (PSD) consisting of dual-chevron MCPs (7) and a delay-line (DL) anode (8). On their way to the PSD, the electrons pass through the helical RF deflector (6) consisting of half period, helical electrodes [6] and a 500 MHz RF power source. The electrons are multiplied by a factor $\sim 10^6$ in the MCP system and the resulting electron cloud hits the DL anode (type DLD40 [7]), producing position-sensitive signals with rise times of a few ns. In this configuration the RFT effectively operates as a Radio Frequency Photomultiplier Tube (RFPMT) [8, 9], capable of timing the arrival of single photons.

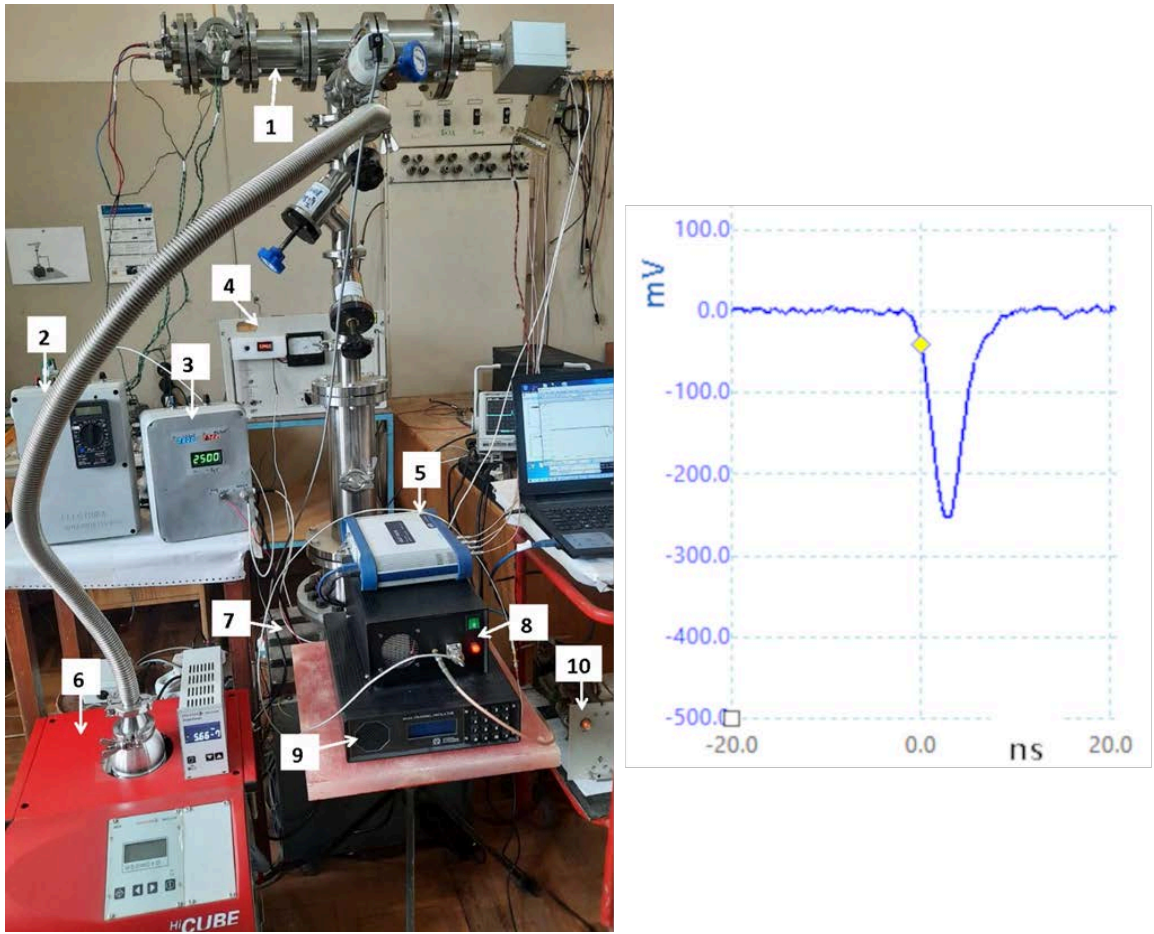

*Figure 2: Left: the RF Timer and test setup. 1-RF Timer prototype; 2-power supply for electrostatic lens, 3-HV power supply for accelerating electrode and MCP detector; 4-HV power Titanium vacuum pump; 5-PICOSCOPE; 6-Pfeiffer vacuum pump; 7-"Titanium" vacuum pump; 8-R ied signal from the DL anode.F amplifier; 9-RF power source; 10-power supply for signal amplifiers. Right: typical amplif*

A photograph of the test experimental setup is presented in Fig. 2 (left) and a typical amplified signal from the DL anode is shown in Fig. 2 (right). Pickup on the anode, induced by the RF power is negligible and does not affect the reconstruction of the image (time) of the RF scanned electrons.

## 3. Measurement of the Timing Resolution at AREAL

The timing resolution of the prototype RFPMT has been measured at the CANDLE, AREAL laser facility [10] which provided 258 nm (4.8 eV) photon bunches (0.45 ps FWHM) phase locked to a 500 MHz oscillator, at a repetition rate of 100 Hz. These were directed to the Tantalum disc cathode and the 500 MHz RF was used to power the RF deflector. The sinusoidal RF signal from the AREAL master oscillator also provided a time reference. Electrons produced by the incident photon pulses of the laser are circularly scanned on to the DL anode, giving *X* and *Y* coordinates. A transform to polar coordinates yields the radius *R* and phase $\varphi$, which is proportional to time ($2\pi \equiv 1/500$ MHz = 2ns). As the laser pulse length is short, all photoelectrons effectively have the same phase and a spot on the scanning circle is obtained. The spread in phase of these points represents the overall time resolution of the system, which includes factors related to the laser, the laser and RF oscillator synchronization device and the intrinsic time resolution of the RFPMT.

Results with four different fixed phases are displayed in Fig. 3, where the equivalent time difference between these phases is about 100 ps. The point in the center of Fig. 3 (left) is a 2D image of the 2.5 keV electrons, obtained when the RF is OFF. The circle is an image of the scanned electrons, when the 500 MHz RF is ON, but not synchronized with the laser and the color spots on the circle correspond to phase distributions of RF-synchronized photoelectrons for the four different fixed phases. The distributions of these phases are shown in 1D in Fig. 3 (middle) and show that the time resolution is ~10 ps.

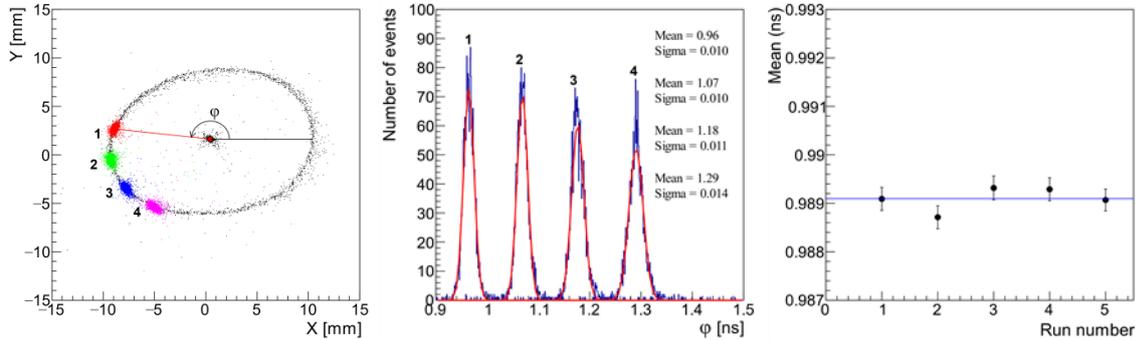

*Figure 3: Left: 2D images of anode hit positions. Middle: distributions of phases. Right: timing stability measurement.*

An analysis of the various factors which contribute to the time resolution has identified chromatic aberration due to the emitted PE's initial energy spread $\Delta\varepsilon$ as the largest contributor. The time resolution for $\Delta\varepsilon = 1$ eV, simulated using the SIMION software package amounts to ~8 ps. Other significant contributions come from the intrinsic size of the electron beam spot and the position resolution of the DL anode and amount to ~6.5 ps. In quadrature they amount to 10.3 ps, in good agreement with the results of Fig. 3 (middle).

The time stability was investigated in a series of measurements at a fixed phase, carried out at ~10 min intervals. The mean values of the sequentially measured phase distributions is shown in Fig. 3 (right), demonstrating that the time stability of the RF PMT over a period of ~1 hour is about 0.5 ps, FWHM.

The prototype RFPMT has a single helical deflector and scans incident electrons on to a circle and is under development to improve the time resolution. At 500 MHz RF power frequency the period of pico-time measurement is 2 ns. This can be extended by using multiple deflectors tuned to different frequencies. For example, spiral scanning with 2 helical deflectors, powered at slightly different frequencies, will offer periods of some tens of ns [11]. The more complicated scanning patterns require a more sophisticated PSD and developments employing TimePix3 are being investigated.

## 4. RFT and RFPMT Applications

The first application of the RFT (lifetime measurements of quantum states of nanostructures) is under construction at the AANL and is described below. This is followed by summaries of potential RFPMT applications: Positron Emission Tomography (PET), DOT, STED and Förster Resonance Energy Transfer (FRET) nanoscopy.

### 4.1 Nanoscience: The study of quantum states with ps lifetimes

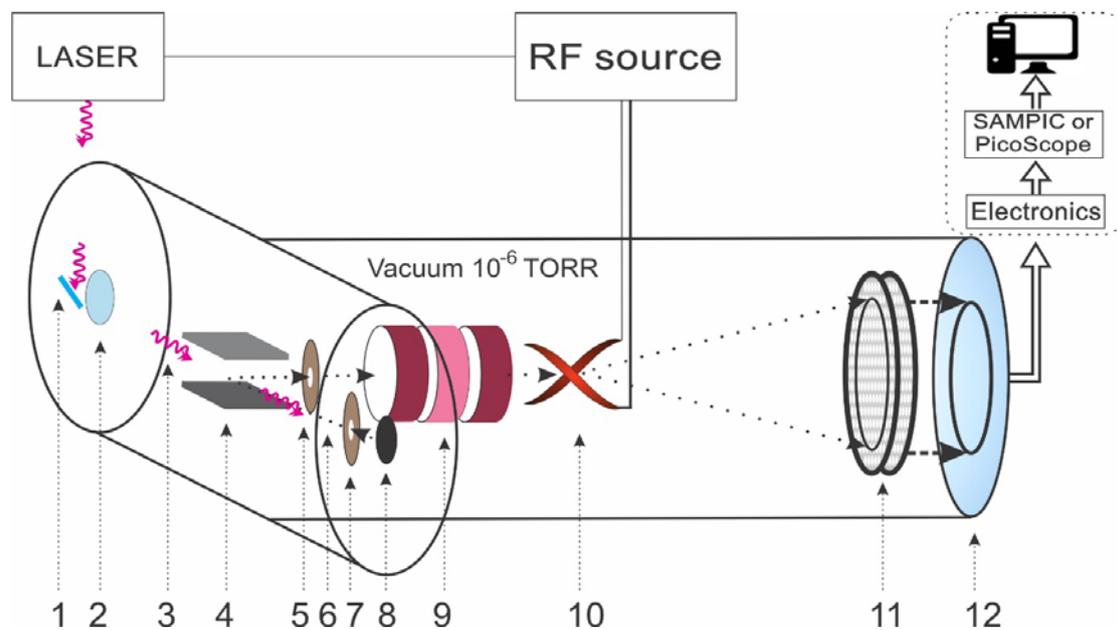

*Figure 3: Schematic layout of the device. (1) mirror; (2) quartz-glass window; (3) incident photons; (4) permanent magnet; (5) collimator; (6) photoelectron; (7) electron transparent electrode; (8) graphene sample; (9) electrostatic lens; (10) RF deflector; (11) MCP detector; (12) position sensitive anode.*

The lifetimes of quantum states in nanostructures after photoexcitation dictates the practicality of these materials in many applications such as solar energy conversion, surface chemistry, photonics and optoelectronics. Because of many attractive properties arising from its unique band-structure, graphene has been considered as one of the most promising materials for optoelectronic devices. Theoretically, the lifetime of hot carriers in graphene should be quite long: hundreds of picoseconds to a few nanoseconds [12]. However, experimental measurements in the past decade, employing various pump-probe methods [13-25], have always measured few ps decay times. The fast decay of hot carriers (similar to that of a metal) and the very small bandgap suggest that graphene might not be ideal for real applications.

Recently, it was shown [26] that at very low irradiation levels, there exists a significant slow-decay process in graphene at room temperature, which is two orders of magnitude slower than previous results [13-25]. This was attributed to the excitation of image potential states (IPS) with long lifetimes. The employed laser fluence was very low at ~10 µJ/cm$^2$ and at such low fluencies many experimental techniques become non-applicable, due to the extremely low signal level. However, a new time-of-flight (TOF) angle-resolved photoemission spectroscopy (ARPES) apparatus has been developed to achieve a high detection efficiency, thus enabling measurements without employing a conventional pump-probe setup. It has been proposed [26] that the hot carriers in graphene can be excited to IPS at low laser fluencies. Thus, it will be interesting to investigate how important these IPS electrons are at even lower laser fluencies and whether these carriers can be effectively extracted to implement opto-electric devices. The apparatus (Fig. 4) consisting of the current prototype RFT, and an RF synchronized laser is under construction at the AANL. It will perform a few-picosecond-precision measurement of electrons ejected from a graphene sample. In addition to high time resolution, it will have high detection efficiency, which will allow the use of lower laser fluencies.

## 4.2 Time-OF- Flight Positron-Emission-Tomography

The positron emission tomography (PET) is a powerful technique used in medicine and medical research to image the molecular process in vivo. It provides a highest sensitivity at a pico-molar level for the imaging modalities together with a good spatial resolution. An important advance of the PET technique the use of time-of-flight (TOF) ([27, 28] and references therein). For TOF one measures the detection time of each registered gamma-quantum and uses the time difference between two coincident gammas to localize the emission point along the line-of-response. The additional information from TOF significantly improves the quality of the reconstruction image. The best TOF performance from a commercial system (Siemens Biograph Vision) yielded a coincidence time resolution (CRT) of 210 ps [29], but the real breakthrough is expected when a CRT of 10 ps is achieved [30]. A (sub)-10 ps capability increases by a factor ≥16 the effective PET sensitivity, compared to non-TOF whole body PET. Such an improvement will result in a reduction of the radiation doses to extremely low levels and open the road to a scanner with high spatial resolution. It will also accelerate the TOF image reconstruction by a factor 200 or more [30]. The RFPMT [1], is an excellent candidate to achieve such performance when combined with a Cherenkov radiator such as crystalline PbF$_2$ [31, 32] or TMBi liquid [33]. A schematic of a large-diameter photocathode RFPMT, suitable for application in TOF-PET is presented in Fig. 5.

An incident gamma produces Cherenkov photons in the radiator which strike the photocathode producing photoelectrons. These are accelerated and focused in the "spherical-capacitor" region and at the focal point they pass through a transmission dynode producing secondary electrons (SE). Low energy SE produced on the rear side of the transmission dynode are accelerated through the electron transparent electrode and focused by the electrostatic lens, before passing through the RF deflector. The scanned SE are then detected on the PSD as shown in Fig.1.

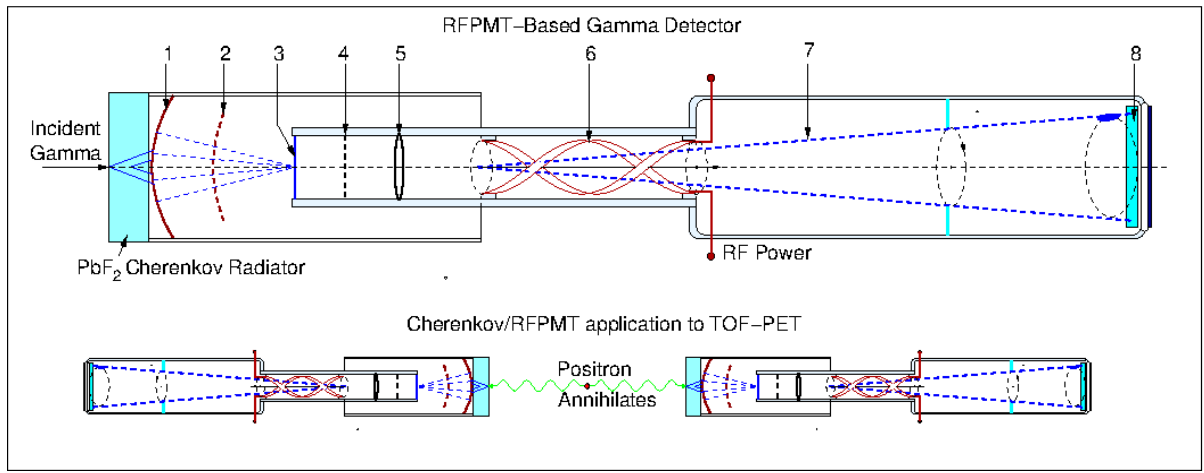

*Figure 4: Schematic of the Cherenkov/RFPMT gamma detector and its application to TOF-PET: 1-large-diameter spherical photocathode; 2-electron transparent electrode; 3-transmission dynode; 4-accelerating electrode; 5-electrostatic lens; 6-RF deflector; 7-RF deflected SE; 8- position-sensitive detector.*

The Cherenkov/RFPMT detector [34] is projected to be capable of ~10 ps time resolution for an individual module, leading to ~15 ps for the CTR. The same order of CRT could also be achieved by using fast scintillator and an RFPMT with a spiral scanning system and pixelated anode [35].

## 4.3 Time-of-Flight Diffuse Optical Tomography (TOF- DOT)

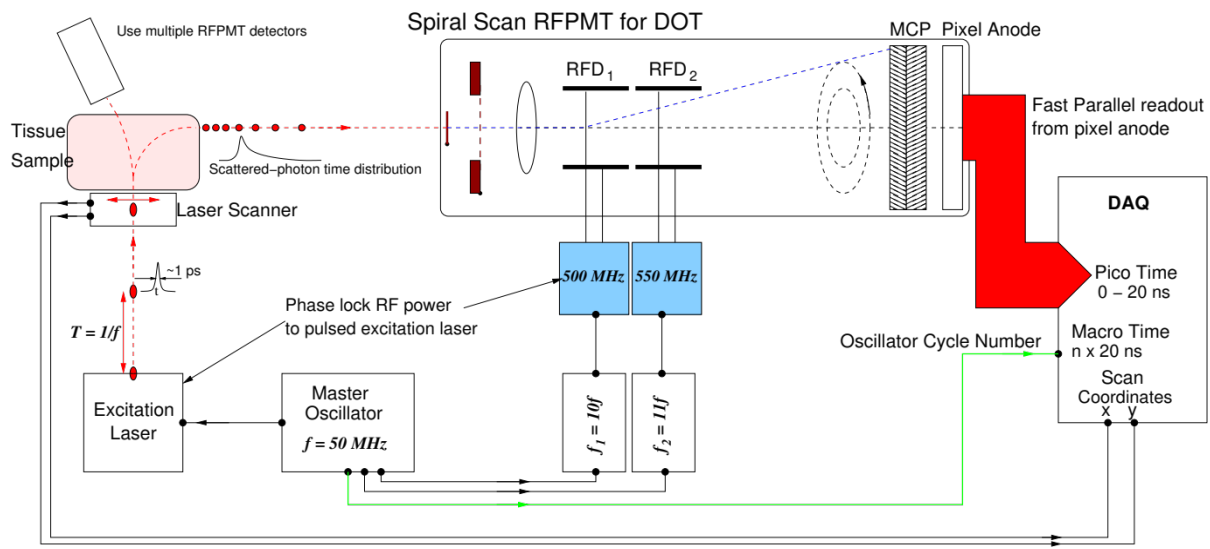

*Figure 5: Conceptual diagram of the use of a spiral-scan RFPMT in DOT.*

Optical techniques have particular therapeutic potential in the medical field, because visible and infrared (IR) light is, unlike x rays, non-ionizing and gives significantly better spatial resolution than ultrasound. Moreover, different types of tissues interact differently with visible light. One of the most useful features is the difference in the absorption coefficient at 700 nm of oxidized and non-oxidized hemoglobin, which can be utilized not only in identifying an activated metabolism but also in tumor detection [36, 37]. In addition, the scattering coefficient depends sensitively on the cells' organelles, such as the nucleus and mitochondria. Therefore, visible, and near-IR light is extremely useful for diagnosing different anomalies in biological tissue. Optical imaging in general and optical mammography are therefore among the holy grails of clinical imaging [38].

Photons travelling in scattering media such as biological tissue, can be divided into three broad categories [4, 5]:

- ballistic photons, where there is no interaction with the medium and the photons propagate straight through with a coherent wavefront;
- snake photons which are weakly scattered, arrive immediately after ballistic photons and maintain some coherence since there are only minor deviations in direction;
- diffuse photons which have scattered many times and have a trajectory that no longer correlates with their initial propagation direction, leading to an incoherent wavefront.

Ballistic and snake photons find many uses for imaging through scattering media [4, 5, 39-42].

In 1990, it was demonstrated by Yoo and Alfano [43] that a streak-camera with a time resolution of 8 ps could differentiate between the ballistic and diffusive components of light in a diffusive medium. Subsequently a timing technique with ~10 ps resolution, based on an ultrafast optical shutter [44], was used to create a high-resolution image through a 3.5-mm-thick section of human tissue. Since then, the technology has improved and has become an important tool in the imaging of diffusive tissue [45–57]. It has recently been shown that spatially resolved TOF information of the photons transmitted through semi-opaque tissue, analyzed with an advanced computational imaging technique, can detect hidden objects at cm depths with a resolution of a few mm. This can be improved by increasing both the spatial and temporal resolution of the detected photons [58, 59].

The ballistic and diffusive components of light can also be differentiated by placing restrictions on the collection angles of the detected photons [48, 56, 60, 61]. Recently it was shown that a 1-mm-wide object immersed in 1 to 2 cm of a biological tissue medium can be resolved [38]. Therefore, a combination of timing and angle information allows all photons, i. e. ballistic, snake and diffuse, to be used for imaging of semi-opaque human tissue.

The RFPMTs is ideally suited for these purposes. A potential TOF-DOT application of the RFPMT is displayed in Fig. 6 and much of the setup is similar to the RF-STED application described in Sec. 4.4. Laser pulses are directed at the tissue under study, via a system to scan the incident x-y position. As they propagate through the layers of tissue, the laser pulses are attenuated, and their time profiles broadened. TCSPC, employing multiple RFPMTs to detect the scattered photons at different positions and angles, then gives measurements of the time distributions of photons after they leave the sample via different propagation paths. The measured time distributions $d(t) = I(t) \otimes d_{inst}(t)$ are a convolution of the actual photon time $I(t)$ and the instrumental response $d_{inst}(t)$. Using the spiral-scan RFPMT one could resolve the photon arrival time to ~1 ps, for multiple photons spaced at intervals as short as ~ 1 ps, over a time range of several tens of ns. This level of time resolution and freedom from dead-time distortion leads an RFPMT measurement to have vastly reduced instrumental distortion, compared to more conventional PMT or SPAD devices. With a more accurate measure of $I(t)$, the absorption and scattering properties of the tissue may be determined more accurately, leading to improved quality of imaging.

## 4.4 A Potential STED Application.

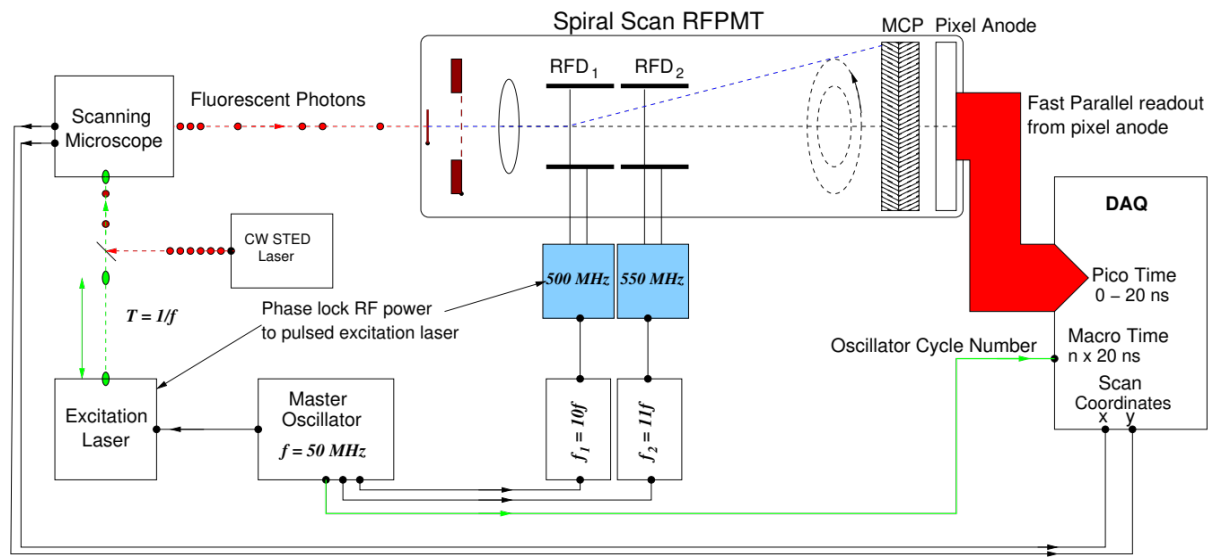

*Figure 6: Schematic of the RF-STED nanoscope for SPLIT-STED microscopy.*

STED microscopy [3, 62] relies on detecting fluorescence after a dye in the sample of interest is excited by a sub-ps pulsed laser, tuned to the absorption spectrum of the dye. Photons from a second laser, red-shifted with respect to the excitation pulse, follow the excitation pulse and constitute the STED beam. This interacts with molecules in the excitation region, effectively quenching them to the ground state by stimulated emission and inhibiting the natural fluorescence. By configuring the spatial extent of the depletion pulse to a doughnut shape, with zero central intensity, only the central spot of the primary fluorescence distribution remains. The extent of the sub-diffractive central spot can be reduced by increasing the intensity of the depletion pulse and by scanning the aligned primary and depletion lasers across the sample, a high-resolution image can be constructed.

Spontaneous de-excitation, with a time constant $\tau \sim$ few ns also takes place, so that the STED process must occur within this time frame. If the arrival time of fluorescent photons, with respect to the excitation laser pulse, is selected or measured precisely, then excellent sub-diffraction resolution can be obtained with reduced peak STED beam intensities. In effect the emission point of the fluorescent photon, with respect to the center of the excitation laser, parametrized in terms of a Point Spread Function (PSF), is correlated to the emission time of that photon. Early photons tend to correspond to the STED maximum of the doughnut, while later photons tend to originate from the central region. Differentiating the arrival times requires a photon detector with highly precise timing resolution.

The RFPMT, with ps resolution for single photons and freedom from dead-time distortion, can map very precisely the arrival times of all fluorescent photons, which will be a convolution of natural-decay and STED-induced time distributions as well as uncorrelated background. This makes it ideal for so called Separation-of-Photons-by-Lifetime-Tuning-STED (SPLIT-STED) microscopy [63], which can achieve using a phasor-type analysis [64] of the measured time distribution. This more sophisticated separation of decay-time distributions will lead to improved image quality, given the high-precision, undistorted time distributions measured by the RFPMT.

Fig. 7 depicts an RFPMT based SPLIT-STED (RF-STED) microscope. In this case a 50 MHz oscillator drives the pulsed excitation laser beam and generates a 500 MHz signal suitable to drive the deflector of the RFPMT synchronously with the excitation beam.

The RFPMT shown in Fig. 7 has a pixelated anode and a dual-deflector spiral scanning system. Picosecond timing in the range 0-20 ns would be obtained from the anode pixel information, while "macroscopic" (>20 ns) time is obtained from the oscillator cycle number. Rather than time-gating the detection of photons, this system would record practically all fluorescence photons associated with an incident laser burst, so that no information is lost. The data acquisition (DAQ) system would also incorporate the position information from the scanning microscope and with fast parallel readout from the pixel detector, extremely high photon counting rates could be achieved. Direct anode pixel readout would be especially useful, were time tagged and time resolved (TTTR) detection of single photons is needed.

**4.5 The FRET Nanoscope**

**Förster Resonance Energy Transfer** (**FRET**) is a non-radiative mechanism of energy transfer between two light-sensitive molecules (chromophores). An excited donor chromophore may transfer energy to an acceptor chromophore through dipole–dipole coupling whose efficiency $\varepsilon \propto 1/r^6$, where $r$ is the donor-acceptor distance. This makes FRET extremely sensitive to small changes in distance and FRET microscopy [65, 66] is thus a powerful technique for the study of molecular interactions within living cells.

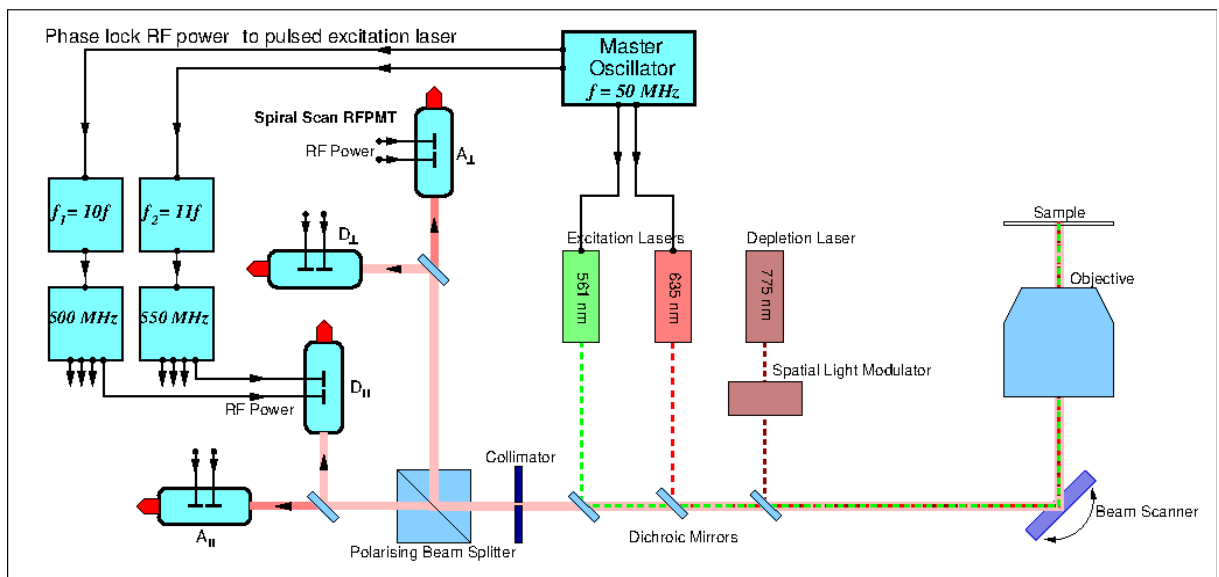

*Figure 7: Schematic of the RFPMT applied to the FRET/STED technique.*

Fluorescence nanoscopy techniques such as STED are not capable of resolving distances at the molecular level. However, a newly proposed technique, FRET nanoscopy [67], combines colocalization STED microscopy with multiparameter FRET spectroscopy. This produces a (sub)nm position resolution of donor and acceptor dyes of FRET pairs. The combined information from FRET (isotropic 3D distance information) and colocalization STED (projected distance to the image plane) enables a reconstruction of the 3D orientation of a molecular structure.

Fig. 8 shows a schematic of this technique, where RFPMTs replace the SPAD photon detectors commonly used in microscopy. With a RFPMT, time measurement in the STED microscopy and FRET nanoscopy can be realized with a resolution better than 10 ps. Lifetimes can be measured with a precision better than 0.5 ps, FWHM.

## 5. Summary

This paper describes the RFPMT and its current and future applications. It is based on a helical deflector, which performs circular or elliptical sweeps of keV electrons by means of a 500 MHz radio frequency electromagnetic field. By converting the time of arrival of incident electrons to a hit position on a circle or ellipse (the locus of scanned electrons determined by the period of helical deflector, this device achieves a timing resolution of 10 ps. This is mainly due to the technical parameters of the prototype tube and can certainly be improved.

Already the RFPMT will be employed for a measurement of the ps lifetimes of quantum states of nanostructures and after further development, has potential applications in several bio-medical fields. Applications in TOF-PET, TOF-DOT, STED and FRET nanoscopy are outlined.

## Acknowledgements

This work was partially supported by the International Science and Technology Center (ISTC) in the framework of scientific project A-2390, the Science Committee of the Republic of Armenia (Grants: 21T-2J133, 20TTCG-1C011,18Ap_2b05 and 21APP-2B012), the ARPA Institute, the ANSEF (Grant hepex-4954), the UK Science and Technology Facilities Council (Grants: ST/V00106X/1, ST/S00467X/1), the JSPS KAKENHI (Grants: 18H05459, 17H01121).